\documentclass[%
 reprint,
superscriptaddress,
 amsmath,amssymb,
 aps,
prb,
floatfix,
]{revtex4-2}
\usepackage{graphicx}
\usepackage{dcolumn}
\usepackage{bm}
\usepackage[utf8]{inputenc}
\usepackage[T1]{fontenc}
\usepackage{mathptmx}
\usepackage{etoolbox}
\usepackage{calc}
\usepackage{comment}
\usepackage{graphicx}
\usepackage[shortlabels]{enumitem}
\usepackage{amsmath}
\usepackage{amsfonts}
\usepackage{amssymb}
\usepackage{float}
\usepackage{xcolor}
\usepackage{spalign}
\usepackage{natbib}
\usepackage{systeme}
\usepackage{units}
\usepackage{xcolor}
\usepackage{bm}

\begin{document}
\title{Raman Scattering with infrared excitation resonant with MoSe$_{2}$ indirect band gap}

%\begin{comment}
\author{Simone Sotgiu}
\email{Corresponding author: simone.sotgiu@uniroma1.it}
\affiliation{Department of Physics, Sapienza University of Rome, Piazzale Aldo Moro 5, 00185 Roma, Italy}

\author{Tommaso Venanzi}
\affiliation{Department of Physics, Sapienza University of Rome, Piazzale Aldo Moro 5, 00185 Roma, Italy}

\author{Francesco Macheda}
\affiliation{Istituto Italiano di Tecnologia, Graphene Labs, Via Morego 30, I-16163 Genova, Italy}
%\affiliation{Istituto Italiano di Tecnologia, Center for Life Nano and Neuroscience, Viale Regina Elena
%291, I-00161 Rome, Italy}

\author{Elena Stellino}
\affiliation{Department of Physics and Geology, University of Perugia, via Alessandro Pascoli, Perugia, Italy}

\author{Michele Ortolani}
\affiliation{Department of Physics, Sapienza University of Rome, Piazzale Aldo Moro 5, 00185 Roma, Italy}

\author{Paolo Postorino}
\affiliation{Department of Physics, Sapienza University of Rome, Piazzale Aldo Moro 5, 00185 Roma, Italy}

\author{Leonetta Baldassarre}
\email{Corresponding author: leonetta.baldassarre@uniroma1.it}
\affiliation{Department of Physics, Sapienza University of Rome, Piazzale Aldo Moro 5, 00185 Roma, Italy}
%\end{comment}

\date{\today}
%\section*{Abstract}
\begin{abstract}
 Resonance Raman scattering, which probes electrons, phonons and their interplay in crystals, is extensively used in two-dimensional materials. Here we investigate Raman modes in MoSe$_2$ at different laser excitation energies from 2.33 eV down to the near infrared 1.16 eV. The Raman spectrum at 1.16 eV excitation energy shows that the intensity of high-order modes is strongly enhanced if compared to the  first-order phonon modes intensity due to resonance effects with the MoSe$_2$ indirect band gap. By comparing the experimental results with the two-phonon density of states calculated with density functional theory, we show that the high-order modes originate mostly from two-phonon modes with opposite momenta. In particular, we identify the momenta of the phonon modes that couple strongly with the electrons to produce the resonance process at 1.16 eV, while we verify that at 2.33 eV the two-phonon modes lineshape compares well with the two-phonon density of state calculated over the entire Brillouin Zone. We also show that, by lowering the crystal temperature, we actively suppress the intensity of the resonant two-phonon modes and we interpret this as the result of the  increase of the indirect band gap at low temperature that moves our excitation energy out of the resonance condition.  \end{abstract}
%\pacs{}
\maketitle

Raman scattering is a widely used spectroscopic technique to study the vibrational and electronic excitations in solids. When the incoming laser energy matches a real electronic transition, in a so-called resonance effect,  an enhancement of the Raman cross-section occurs with respect to non-resonant processes, leading to a higher visibility of otherwise hidden modes and of high-order Raman modes, providing information on electronic transitions, phonon-dispersion and electron-phonon interaction  \cite{cardona2006light,bruesch2012phonons,carvalho2020resonance,lee2018resonance,placidi2015multiwavelength}.

Resonance Raman spectroscopy has been extensively employed to study graphene and other two-dimensional materials as it provides information on their vibrational properties \cite{malard2009raman,ferrari2006raman,caramazza2018first,saito2016raman}, on the presence of defects \cite{eckmann2012probing,wu2017spectroscopic}, and it is used to identify different stacking orders \cite{cong2011raman} and for growth quality check \cite{o2016mapping}.
Among the vast class of two dimensional materials, the semiconductor compounds, such as the transition metal dichalcogenides (TMDs) MX$_2$ with M=Mo,W and X=S,Se,Te, have attracted a particular interest because of their potential applications in opto-electronic devices. The layered crystal structure, which leads to an extreme surface-to-volume ratio in the single layers, makes these materials one of the most promising candidates for the development of flexible and ultraflat opto-electronic devices \cite{mueller2018exciton,wang2012electronics,tonndorf2013photoluminescence,tongay2012thermally,venanzi2021terahertz}. 

 MoSe$_2$ was shown to be attractive for several applications such as  electrochemical energy storage, due to its particular chemical properties and its good conductivity \cite{eftekhari2017molybdenum}, and as near-infrared photodetectors \cite{ko2017high} thanks to its thickness dependent band-gap width (indirect band gap of 1.14 eV for MoSe$_{2}$ bulk and 1.6 eV direct band gap for monolayers) \cite{kumar2015thermoelectric,tonndorf2013photoluminescence}, and its high optical absorption in the near-infrared region \cite{bernardi2013extraordinary}. A deep understanding of the scattering channels, e.g. electron-phonon scattering and of the related charge carrier transport properties \cite{ponce2021first,macheda2020theory,macheda2022fr} is mandatory for the implementation of TMDs in any opto-electronic device. To this end, resonance Raman scattering provides a unique possibility to address the scattering processes in MoSe$_2$ using several excitation energies, from visible to near UV \cite{sekine1980raman, soubelet2016resonance,nam2015excitation,kim2016davydov}. However, to the best of our knowledge, the lowest excitation energy used was 1.58 eV, hence no Raman scattering has been reported so far with incoming photon resonant with the indirect band gap of MoSe$_{2}$, meaning that no study of the interaction of electrons at the bottom of the conduction band with the phonons has been performed up to now.

In this work, we present a resonance Raman scattering study of MoSe$_{2}$ crystals measured with an incoming photon energy of 2.33 and 1.96 eV  and 1.16 eV that matches the indirect band-gap transition of MoSe$_2$. For the latter photon excitation energy we observe a great enhancement of the scattering intensity of high-order processes with respect to first-order modes. By comparing the Raman spectra with the two-phonon density of states (2ph-DOS) calculated by Density Functional Theory (DFT), we associate these peaks with two-phonon resonant Raman processes involving the electronic transition at the indirect band gap of the crystal. Our results deepen the understanding of Raman scattering in MoSe$_{2}$, indicating that phonon modes with momenta comparable to that connecting the indirect band-gap points in the Brillouin zone are strongly coupled to electrons. Our results, moreover, demonstrate that the near-infrared excitation radiation can be used as a powerful tool to study small-gap semiconductors obtaining relevant information on the material properties. \cite{mooradian1966first}.
\section{Methods}
Raman spectra at 1.16 eV excitation energy are collected with a Fourier-transform interferometer (Bruker MultiRAM), equipped with a Nd:YAG laser and a nitrogen-cooled Germanium detector (spectral range 5900-11700 cm$^{-1}$), and connected via optical fibers to an optical microscope. The spectra at room temperature are obtained with a 100x objective (Numerical Aperture 0.85), spectral resolution of 1 cm$^{-1}$ and a power density around 2.5 mW/$\mu$m$^{2}$. The temperature-dependent measurements are performed with a He-flow cryostat and a 4x objective. We set 2 cm$^{-1}$ resolution and 0.02 mW/$\mu$m$^{2}$ power density. The spectra at 2.33 eV and 1.96 eV excitation energy are measured with a Horiba HR Evolution microspectrometer equiped with a 100x objective coupled to a Silicon CCD camera. We measure with 0.6 cm$^{-1}$ spectral resolution and power density of 5 and 0.5 mW/$\mu$m$^{2}$ for 2.33 and 1.96 eV respectively. MoSe$_{2}$ single crystals are purchased from \textit{HQ Graphene}. The absorbance spectrum is obtained from a transmission experiment using Bruker FT-IR spectrometer. For DFT calculations we used a GGA-PBE functional with semi-empirical Grimme-D3 van der Waals corrections \cite{grimme2010consistent} to optimize the geometry of the structure, employing a energy cutoff of 125 Ry and a  $\mathbf{k}$-point grid of dimensions $15\times15\times3$. The optimization leads to a lattice parameter of a = 6.266 Bohr and an interlayer distance of 24.665 Bohr, in accordance with \cite{kim2021thickness}. The position of the electronic band gap is dependent on the interlayer distance, which is incorrectly overestimated if the van der Waals corrections are not taken in account. The 2ph-DOS---namely, $DOS_{2\omega}(\epsilon)=\sum_{\mathbf{q}\mu\nu}\delta(\epsilon-\omega_{\mathbf{q}\nu}-\omega_{\mathbf{q}\mu})$, $\omega_{\mathbf{q}\mu,\nu}$ being the phonon frequency---is obtained computing the phonons using Density Functional Perturbation Theory \cite{baroni2001phonons} on a $\mathbf{q}$-points grid of dimensions  $6\times6\times2$, which is then interpolated  on a $768\times768\times1$ grid to evaluate the sums in the expression for  $D_{2\omega}(\epsilon)$. All the calculations have been performed using the QUANTUM ESPRESSO suite \cite{giannozzi2009quantum}.
\section{Results and Discussion}
\begin{figure}
\centering\includegraphics[scale=0.43]{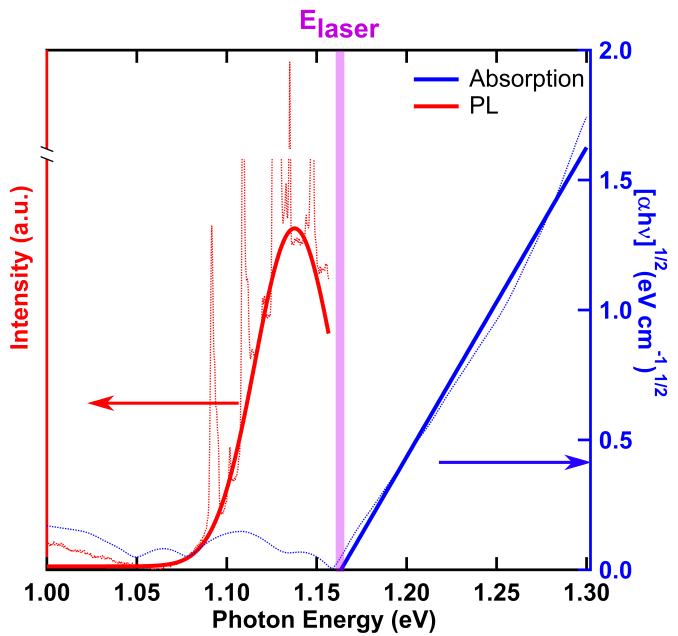}
\caption{Room Temperature PL (red curve) obtained fitting the background of Raman spectrum MoSe$_{2}$ crystal (dotted red curve). The extrapolated gap is E$_{gap}^{PL}=1.137\pm0.001$ eV. Tauc procedure (blue curve)  from absorption experiment (dotted blue curve). The extrapolated gap is E$_{gap}^{Tauc}=1.16\pm0.02$ eV. We have highlighted our 1.16 eV laser energy E$_{laser}$.}
\label{gap_PL_Tauc}
\end{figure}
First of all we determine experimentally the energy of the indirect band gap at room temperature in order to compare it with our laser lines. We measure both the photoluminescence (PL) and the absorbance spectrum and use a Tauc procedure to extract the band-gap width \cite{zanatta2019revisiting,makula2018correctly} (see Figure \ref{gap_PL_Tauc}). The discrepancy between the value obtained from the PL ($E_{gap}^{PL}=1.137 \pm 0.001$ eV) and the value obtained from the absorbance spectrum ($E_{gap}^{Tauc}=1.16 \pm 0.02$ eV) is known as \textit{Stokes Shift} and, in first approximation, can be explained using the Frank-Condon principle \cite{pelant2012luminescence} but it is also modulated by strain \cite{niehues2020strain}, impurities \cite {kanemitsu1998phonon,martin1994theory} and other factors. We remark that a deeper understanding of this phenomenon in MoSe$_2$ is beyond the scope of this article, here we just point out that our IR laser energy well matches the indirect band-gap width.

\noindent
\subsection{Raman spectra as a function of excitation energy: experiment and theory}
\begin{figure}
\centering
\includegraphics[scale=0.31]{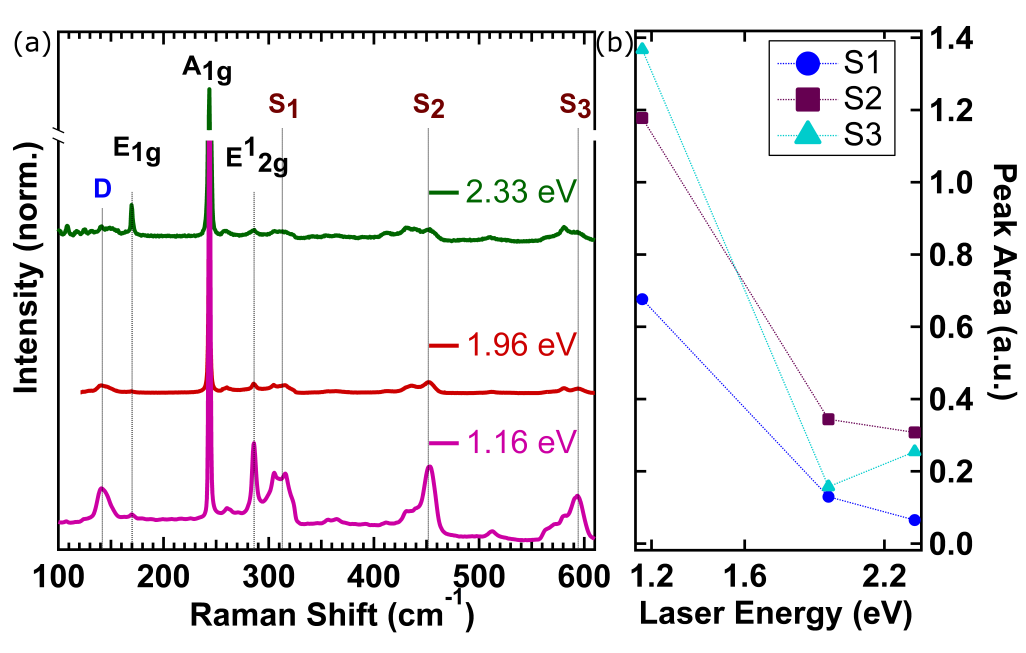}
\caption{(a) Spectra of MoSe$_2$ bulk taken at infrared (1.16 eV) and visible (1.96 and 2.33 eV) energies. We can observe the presence of the first order Raman peaks (labelled in black) and high order modes due to difference (labelled in blue) or summation (labelled in red) processes (see main text for their definition). The high order Raman peaks  in the spectrum taken at 1.16 eV are strongly enhanced. The E$_{1g}$ peak becomes visible as the excitation energy approaches the energy of the C-exciton (2.6 eV). All the spectra are normalized at the intensity of the A$_{1g}$ mode and vertically shifted for sake of clarity. (b) Integrated Area of the S- Raman features, normalized to the Area of the A$_{1g}$ peak for the different laser energies used. One can clearly see the intensity enhancement of the S- modes with respect to first-order Raman peaks.}
\label{Vis_IR_comp_edited}
\end{figure}

2H-MoSe$_2$ belongs to the space-group symmetry $D^{4}_{6h}$ and it has 12 modes of lattice vibrations at the center of the First Brillouin Zone (FBZ) \cite{sekine1980raman}, namely
\begin{equation}
\Gamma=A_{1g}+2A_{2u}+B_{1u}+2B_{2g}+E_{1g}+2E_{1u}+2E_{2g}+E_{2u}
\end{equation}
where E$_{1g}$, A$_{1g}$, E$_{2g}^{1}$ and E$^{2}_{2g}$ are the Raman-active modes.\\
Figure \ref{Vis_IR_comp_edited}a shows the Raman spectra obtained at three different excitation energies. We identify three of the four first order peaks (E$_{1g}$ at 170 cm$^{-1}$, A$_{1g}$ at 243.5 cm$^{-1}$ and E$_{2g}^{1}$ at 285 cm$^{-1}$). The E$^{2}_{2g}$ is at 32 cm$^{-1}$ and below the spectral range of our FT-Raman setup. It is worth noticing that the E$_{1g}$ mode is usually forbidden in back-scattering geometry, but it is visible at 2.33 eV probably because the laser excitation energy approaches the energy of the C-exciton (about 2.6 eV) \citep{nam2015excitation, soubelet2016resonance}.\\

\begin{figure*}[htp]
\centering
\includegraphics[scale=0.13]{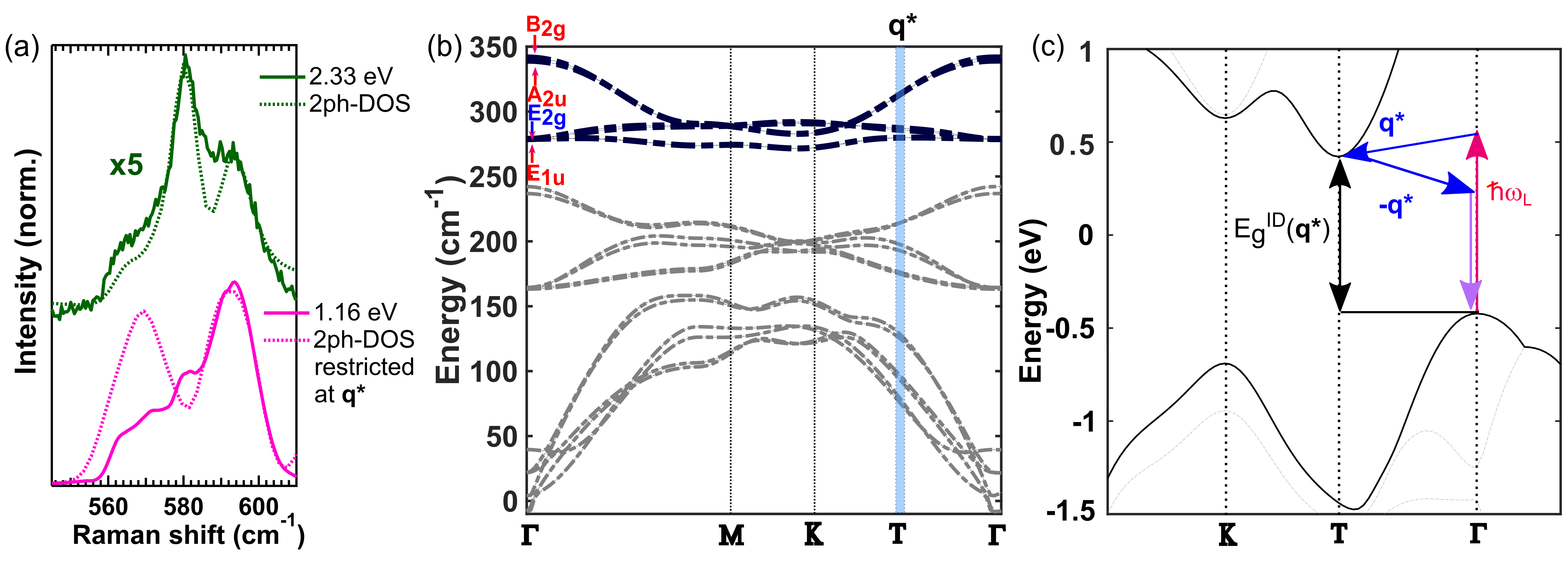}
 \caption{(a) Raman spectra of S$_{3}$ modes compared with the 2ph-DOS. Besides the enhancement of the Raman scattering by using near-infrared excitation, the lineshape of the peaks varies by changing the excitation laser energy revealing a resonant process. The spectra are normalized to the intensity of the A$_{1g}$ mode. (b) Phonon dispersion of bulk MoSe$_2$ restricted to the FBZ along the \boldsymbol{$\Gamma$}-\textbf{M}-\textbf{K}-\boldsymbol{$\Gamma$} line. The $\textbf{T}$ point, i.e. the minimum of the conduction band, is highlighted, together with the phonon wavevector which allows the indirect electronic transition. The S$_3$ feature is then obtained as the direct sum of phonon branches highlighted in black, its energy being hence the sum of the energy of the two branches restricted to a neighbourhood of \textbf{T}. In red are denoted the IR active modes, in blue those Raman active. (c) Scheme of the resonant process described in this work: two phonons with opposite momenta can allow the indirect transition making the Raman process resonant. Note that the value of the energy gap is slightly underestimated by our DFT results, which is a typical feature of these calculations \cite{tongay2012thermally}. }
\label{600_phonon}
\end{figure*}

We can also identify four strongly structured Raman modes, labelled as \textit{D}, \textit{$S_{1}$}, \textit{$S_{2}$} and \textit{$S_{3}$} in Figure \ref{Vis_IR_comp_edited}a, that cannot be ascribed to first order Raman scattering peaks. One can note that these modes are strongly enhanced in the spectrum taken at 1.16 eV, which is a signature of resonance processes (Figure \ref{Vis_IR_comp_edited}b).
Moreover, by looking at the S$_{3}$ mode (see Figure \ref{600_phonon}a) one can note that the overall lineshape is modified.

We first verify whether these are second order Raman modes.
%\subsection{Comparison of experimental data and theoretical calculations}
 A Stokes second-order Raman process is related to the creation of two phonons or to the creation of one phonon and the destruction of another \cite{potts1973temperature}. The resulting Raman shift will be $\hbar(\omega_{1}\pm \omega_{2})$, where $\hbar\omega_{1,2}$ indicates the energy of the phonons and the sign is associated to creation and destruction of the phonon. One should recall that in a two-phonon process, the momentum conservation is achieved by imposing that the two phonons have opposite momentum and therefore no scattering with defects is required. Thus phonons in the entire FBZ are observable with Raman spectroscopy as long as the momentum is conserved and the non-resonant second-order Raman spectra map the two-phonon density of states, as shown for example in diamond \cite{windl1993second}. 
 
We calculate the phonon dispersion (Figure \ref{600_phonon}b) and the 2ph-DOS (see methods), in order to assign the phonon branches that could be involved in two-phonon scattering processes. We compare the spectrum taken with 2.33 eV photon energy with the 2ph-DOS integrated over the entire FBZ. Even without considering the matrix elements for electron-phonon and electron-light interaction, we find a striking similarity for both the energy position and the line shape between the theoretical calculations and the experimental S$_{3}$ features (see Figure \ref{600_phonon}a upper curves). This excellent lineshape agreement strongly suggests that at least the S$_{3}$ mode is due to two-phonon processes only and that using an excitation energy of 2.33 eV we do not select resonantly phonons with a specific momentum but phonon from the entire FBZ contribute equally.
Therefore, we assign the S$_{3}$ mode to the sum of two phonons with opposite momenta from the B$_{2g}$, A$_{2u}$, E$_{2g}$ and E$_{1u}$ optical phononic branches (see Figure \ref{600_phonon}b).
The second-order  S$_{3}$ mode in Figure \ref{600_phonon}a, taken at 1.16 eV, shows a different lineshape. This demonstrates that there are resonant processes occurring and that not all the regions in the FBZ contribute equally in the Raman cross section. Hence, we ascribe the enhancement and the modification in the lineshape of these high-order modes obtained at 1.16 eV excitation energy to the processes resonant with the indirect band gap, as was seen previously in Si and GaP \cite{klein1974selective}. 
Our DFT calculations (Figure \ref{600_phonon}c) confirm that the maximum of the valence band is located at \boldsymbol{$\Gamma$}, while the minimum of the conduction band lies on a point, which we will call \textbf{T}, which is $\mathbf{T}=\frac{t}{3}\mathbf{b_{1}}+\frac{t}{3}\mathbf{b_{2}}$ being $t$=0.55, $\mathbf{b_{1}}$ and $\mathbf{b_{2}}$ the reciprocal lattice vectors, and it is situated along the \boldsymbol{$\Gamma$}-\textbf{K} line, in accordance with previous works \cite{kumar2015thermoelectric,coehoorn1987electronic,kim2021thickness}.
We can thus depict the resonant process as the following: an electron from the top of the valence band is excited and then scattered by a phonon with a wavevector $\textbf{q*}\sim\mathbf{T}$ which allows the transition to the conduction band minimum. Then, a second phonon with opposite wavevector \textbf{-q*} brings the electron to the starting point of the FBZ and then it can relax radiatively emitting a photon (see Figure \ref{600_phonon}c). 
%The vector \textbf{q*} along the $\Gamma$-K line is obtained by the electronic dispersion as:
%$\textbf{q*}=\frac{T}{3}\textbf{b}_{1}+ \frac{T}{3}\textbf{b}_{2}$, where \textbf{b}$_{1}$ and \textbf{b}$_{2}$ are the in-plane reciprocal lattice basis vectors and T=0.55, in agreement with \cite{kim2021thickness}.\\
The above description can be summarized with the vanishing of the second denominator in the following expression for the intensity of the Raman process \cite{klein1974selective}:%\ref{eq:raman}
\begin{widetext}
\begin{equation} \label{eq:raman}
I\propto \left| \sum_{\textbf{k}} \frac{\left<v\left|\textbf{P}\right|c\right>\left<c\left|H_{ep}^{(1)}\right|c'\right>\left<c'\left|H_{ep}^{(1)}\right|c\right>\left<c\left|\textbf{P}\right|v\right>}{[E_{g}^{D}(\textbf{k})-\hbar\omega_{L}][E_{g}^{ID}(\textbf{k}+\textbf{q*})+\hbar\omega_{\mu}(\textbf{q*})-\hbar\omega_{L}][E_{g}^{D}(\textbf{k})+\hbar\omega_{\mu}(\textbf{q*})+\hbar\omega_{\nu}(\textbf{-q*})-\hbar\omega_{L}]}\right|^{2}
\end{equation}
\end{widetext}
\noindent
where the summation runs over the FBZ wavevectors, $\hbar\omega_{L}$ is the laser energy, $\left<v\left|\textbf{P}\right|c\right>$ is the momentum matrix-element between the conduction band minimum and the valence band maximum at the \boldsymbol{$\Gamma$} point, separated by an energy of $E_{g}^{D}(\textbf{k})$ and $H_{ep}^{1}$ is the electron-phonon Hamiltonian connecting the two states at \textbf{k} and \textbf{k}+\textbf{q*}, separated by the energy of the indirect band-gap $E_{g}^{ID}(\textbf{k}+\textbf{q*})$, $\mu$ and $\nu$ being the phonon branches. As it can be seen from the formula above, the Raman processes are strongly enhanced if they involve phonons with opposite momenta \textbf{q*} and the denominator vanishes, i.e. $[E_{g}^{ID}(\textbf{k}+\textbf{q*})+\hbar\omega_{p}(\textbf{q*})-\hbar\omega_{L}]$=0. We remark that there are other possible resonant scattering pathways that include scattering with one electron and one hole (or with two holes). Since we assume constant matrix elements these processes are all equivalent. 

In order to qualitatively understand which phonons contribute to the resonance process, we restrict the sum for DOS$_{2\omega}(\epsilon)$ to a small neighborhood of \textbf{q*} and compare the result with the spectrum taken at IR excitation energy (Figure \ref{600_phonon}b lower curves).
As it can be easily seen, by restricting the calculation of the 2ph-DOS at the particular wavevector \textbf{q*}, the spectral weight of the modes strongly changes, obtaining a qualitative agreement with the experimental spectrum at 1.16 eV. The 2ph-DOS does not match exactly the spectrum, since a complete knowledge of the matrix elements in eq.(\ref{eq:raman}) would be needed, but this would fall beyond the scope of this paper. The qualitative agreement between calculation and experimental spectrum at 1.16 eV is a clear indication that the $S_{3}$ modes mostly originate from the depicted process. We note that a slight horizontal shift, around 5 cm$^{-1}$, was necessary to match the 2ph-DOS and the experimental spectra, due to possible DFT errors. 
\begin{figure}
\centering
\includegraphics[scale=0.33]{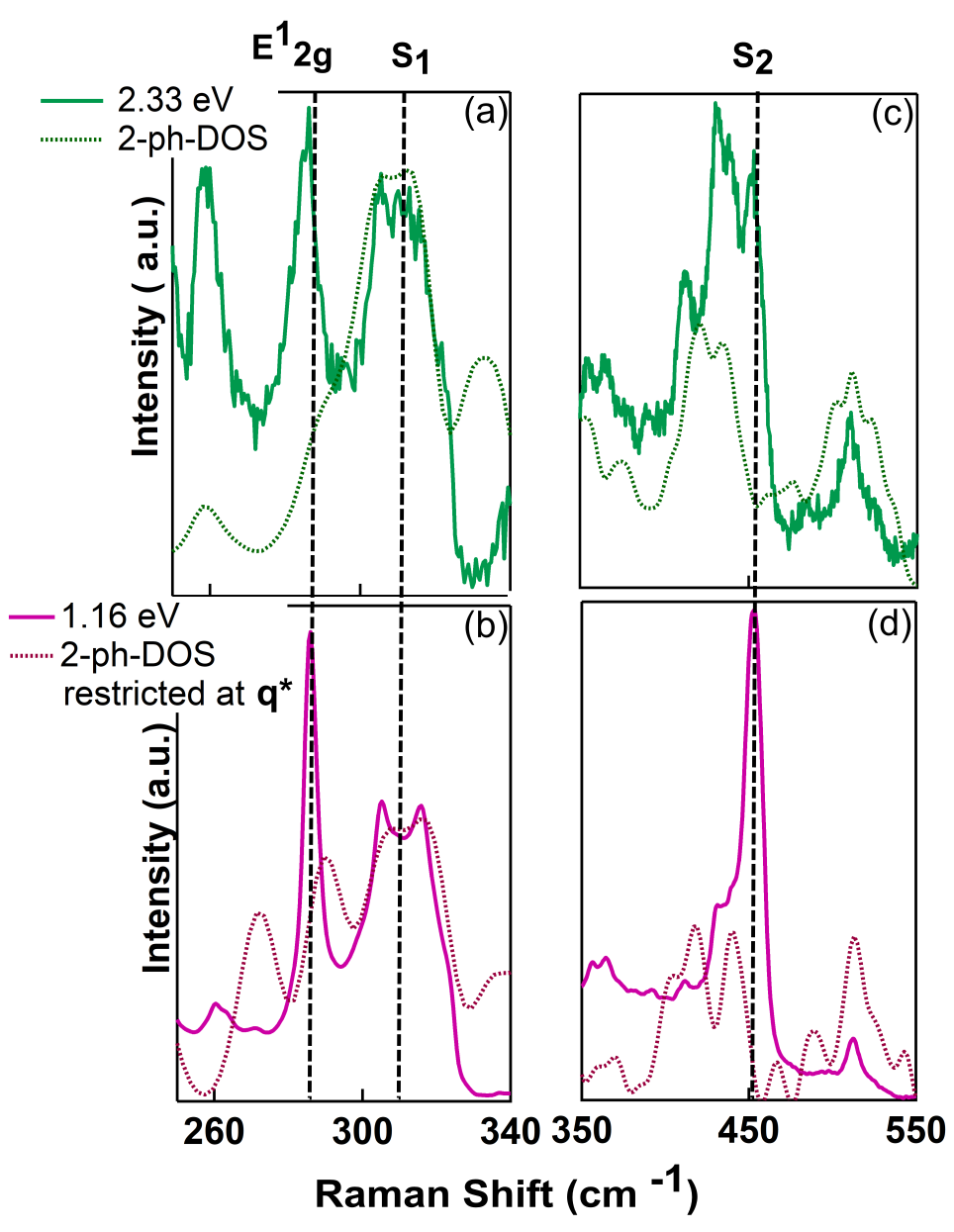}
\caption{(a)-(b) Comparison between the spectrum taken at 2.33 eV and the 2ph-DOS integrated over all the FBZ for different Raman Shift ranges. (c)-(d) Comparison in the same spectral ranges for the spectrum taken at 1.16 eV and the 2ph-DOS restricted at \textbf{q*}.}
\label{RRS_DOS}
\end{figure}
Indeed slight changes to the lattice parameters and anharmonic effects can, for instance, have different effects on phonon frequencies in different zones of the FBZ, therefore resulting in overall shifts that may even be different for the total and restricted 2ph-DOS.\\
In previous studies in literature \cite{soubelet2016resonance,nam2015excitation,chen1974second,golasa2014multiphonon,golasa2014resonant,stacy1985raman,frey1999raman,chakraborty2013layer}, high-order modes assignment have been proposed, mostly linking the high-order peaks to LA modes at the M point, by following the criteria that flat phonon dispersions favor defect-assisted scattering processes.  Here, by comparing our 2ph-DOS to the experimental curves, we can revise such assignment for some of the high-order modes.

The S$_{1}$ modes both for 2.33 eV (see Figure \ref{RRS_DOS}a) and 1.16 eV (see Figure \ref{RRS_DOS}b) excitation laser energy have a similar lineshape to that of the calculated 2ph-DOS. In our calculations all  the possible phonon pairs whose energy summed is 310 cm$^{-1}$ are considered (e.g., A$_{1g}$ + E$_{2g}$ or also B$_{2g}$ + E$_{1g}$ and so on), suggesting that several possible processes build up this peak. Notably, from the 2ph-DOS restricted at \textbf{q*}, a peak around 285 cm$^{-1}$ is evident (see Figure \ref{RRS_DOS}b) and could explain why we observe an intense $E_{2g}^{1}$ peak, with respect to what was measured in previous works \cite{nam2015excitation}. In fact, the $E_{2g}^{1}$ mode was shown to be resonant with the C-exciton ($\sim$ 2.6 eV) and to disappear completely for lower excitation energies \citep{nam2015excitation}. We suppose that in our measurements at 1.16 eV excitation energy, this mode is enhanced by the underlying two-phonon resonant process occurring at the same energy.

The 2ph-DOS does not reproduce the S$_{2}$ modes very well. Besides a strong shift in the absolute energy (see Supplemental Material (SM) \cite{SuppMat}, section A), there is no clear match in the line shapes: a reasonable agreement can be seen between the 2ph-DOS and the experimental spectrum at 2.33 eV, in particular the feature around 520 cm$^{-1}$ is very well reproduced (Figure \ref{RRS_DOS}c), while several differences can be found in the comparison between the restricted 2ph-DOS and the 1.16 eV Raman spectrum (Figure \ref{RRS_DOS}d). Furthermore, we had to add a shift of 20 cm$^{-1}$ to the calculated DOS in order to match the data. This is an indication that, on the one hand, the S$_{2}$ modes could originate by modes with more than two phonons \cite{soubelet2016resonance} that are not considered in our calculation or, on the other hand, that one should implement in the theory the matrix elements to better evaluate the 2ph-DOS or even that the energy of the phonon branches at the edges of the FBZ could be underestimated by our DFT calculations. We note that we could not find any experimental data of the phonon dispersion in literature, i.e. no neutron or x-ray scattering experiments, and we could not compare the DFT calculated phonon dispersion with any data.
We also note that we do not observe any peaks for Raman shifts above 600 cm$^{-1}$, suggesting that modes with order higher than two are not strong and are below our noise level.

In the literature the D-peak has been assigned to a E$_{2g}$- LA difference mode, enhanced by a process resonant with the A-exciton at 1.6 eV \cite{nam2015excitation}. Even if 1.16 eV excitation energy is far from the A-exciton resonance, we still observe an enhancement of the \textit{D} mode around 148 cm$^{-1}$.  The enhancement could originate from a process similar to the one discussed above for the S-modes.
We notice also that similar resonant processes to those discussed in this paper for bulk MoSe$_2$ can be observed in monolayers for higher excitation energies. For example by measuring at 1.96 eV monolayer MoSe$_2$ we find a strong enhancement of S-like modes (see SM \cite{SuppMat}, section B ), however since therein several possible scattering processes can occur, the identification of a single resonant pathway is not easily achievable. 

\subsection{Temperature dependence of Raman peaks}
\begin{figure}
\centering
\includegraphics[scale=0.45]{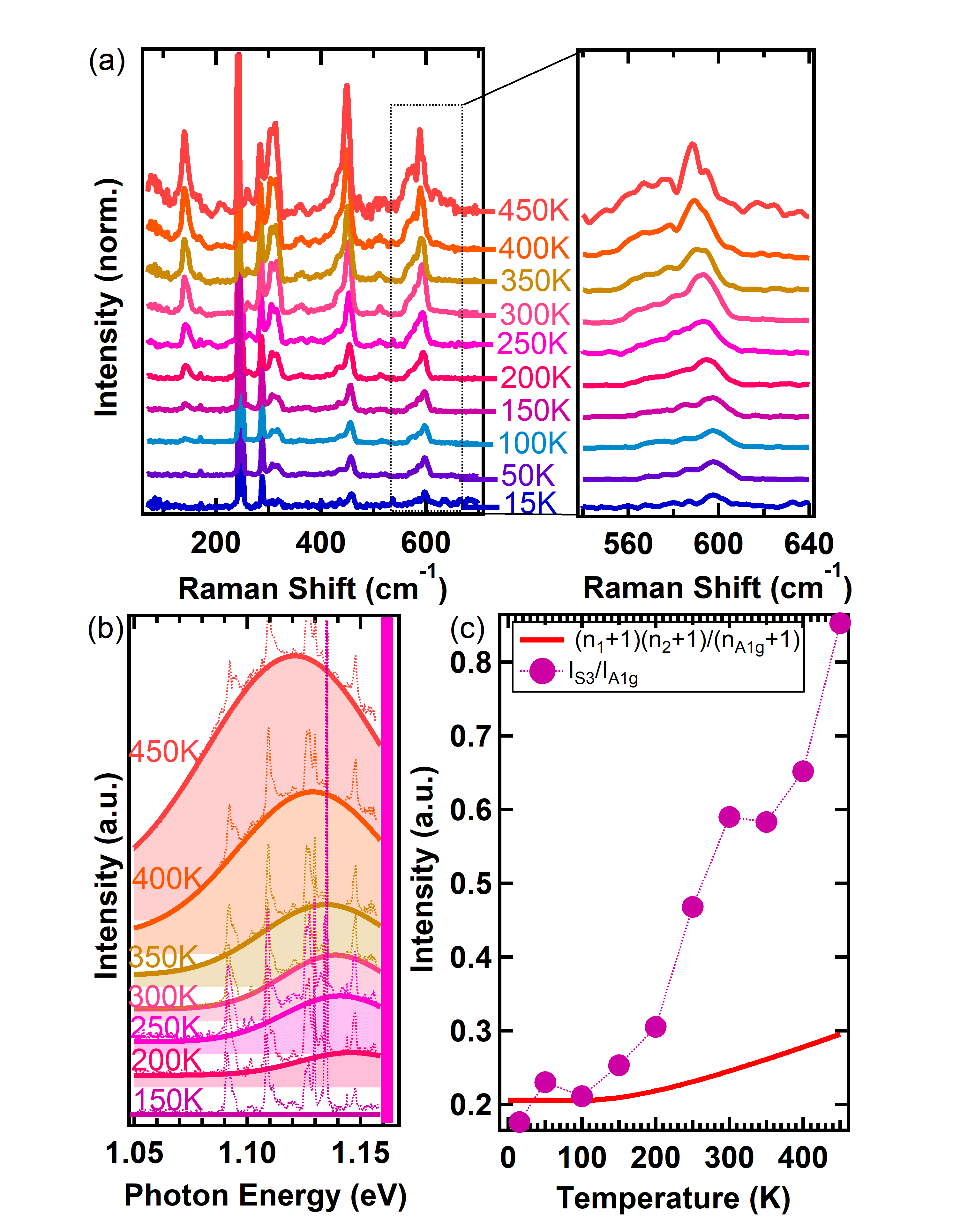}
\caption{(a) Temperature dependent Raman spectra of MoSe$_{2}$ taken at 1.16 eV plotted between 100-700 cm$^{-1}$ and  a zoom of the S$_{3}$ modes. The PL background has been subtracted to the spectra  that are then normalized to the A$_{1g}$ peak. Spectra have been offsetted for sake of clarity. (b) PL peak as extracted from Raman measurements for several temperatures indicated in figure. We observe a PL emission between 200 K and 450 K suggesting that in this temperature range we are able to reach the conduction band minimum through an indirect electronic transition, hence the resonance condition is fulfilled for the second-order Raman process described above. The vertical line represents the laser energy. (c) Comparison between the thermal occupancy given by the BE statistics and the ratio of the S$_{3}$ modes and the A$_{1g}$ peak. We used the values $\hbar\omega_1$=310 cm$^{-1}$ and $\hbar\omega_2$=286 cm$^{-1}$, as explained in the main text. The BE curve has been multiplied for a constant so to match the values of the experimental points at low temperature (i.e. to the average of the I$_{S3}$/I$_{A1g}$ value of the three points below 100 K where there is no PL peak, hence no resonance condition). }
\label{Temp_PL}
\end{figure}
Further proof of the resonant origin of our S- modes could be provided with temperature dependent measurements. Indeed, it was recently shown that the indirect band gap of bulk MoSe$_2$ increases of about 100 meV by lowering the temperature \cite{kopaczek2022temperature}. We thus aim at further demonstrate the resonant origin of the S modes by driving the sample out of resonance with external temperature. In Figure \ref{Temp_PL}a, we report the Raman spectra collected with 1.16 eV excitation energy between 15 K and 450 K. We see that the dominant effect of lowering the temperature is the strong reduction of the intensity of the S- and D- modes. In particular, the spectral feature \textit{D} completely disappears at 15K, while all the other second order modes are only reduced as the temperature is lowered. Figure \ref{Temp_PL}b shows the PL emission between 150 K and 450 K. We note that the data show an increase of the PL intensity and a red-shift of the correspondent band center with temperature, suggesting that in this temperature range the indirect electronic transition can be accomplished with the infrared laser energy used, this leads to the vanishing of the second denominator in eq. (\ref{eq:raman}), meaning that the resonance condition is still fulfilled. What we can notice is that a minor peak at 360 cm$^{-1}$ Raman Shift gains spectral weight by increasing the temperature above 350 K and that the lineshape of S$_{1}$ and S$_{3}$ modes are modified. We can interpret this as a result of the reduction of the indirect bandgap, that thus allows for extra scattering pathways in MoSe$_{2}$. Indeed, only by means of thermal occupation we would not expect a modification of the lineshape.

First of all we consider the temperature dependence of the Raman intensities arising from the Bose-Einstein (BE) statistical occupation. A Stokes process, with an energy shift of $\hbar\omega_1$, has a temperature dependence of [n($\omega_1$,T)+1], where n($\omega_1$,T)=(e$^{\hbar\omega_1/k_{B}T}-1$)$^{-1}$ is the BE factor, T being the temperature of the crystal and k$_B$ the Boltzmann constant. For a two-phonon process, we must consider the product [n($\omega_1$,T)+1][n($\omega_2$,T)+1], where $\hbar\omega_{1}$ and $\hbar\omega_{2}$ are the energies of the two phonons involved \cite{potts1973temperature,mitioglu2014second}. At each temperature, a fitting procedure has been applied to A$_{1g}$ and higher order modes using a multi-gaussian plus a linear function baseline (see SM \cite{SuppMat}, section C, for the fit procedure). We divide the intensity obtained from the fit of the higher-order modes to the corresponding intensity of the A$_{1g}$ peak and we compare the results with the BE ratio [n($\omega_1$,T)+1][n($\omega_2$,T)+1]/[n($\omega_{A_{1g}}$,T)+1].

Figure \ref{Temp_PL}c shows the temperature dependence of the BE statistics for a two-phonon process and the experimental intensity ratios for the S$_{3}$ mode (see SM \cite{SuppMat}, section D for more details about the BE trend). We used the values $\hbar\omega_{1}$=310 cm$^{-1}$ and $\hbar\omega_{2}$=286 cm$^{-1}$ for estimating the theoretical trend. These are intended to be only qualitative values extracted from the phonon dispersion around \textbf{T} for the high energy optical phonon branches (Figure \ref{600_phonon}b) with the constrain that $\hbar\omega_{1}$+$\hbar\omega_{2}$ should be equal to 596 cm$^{-1}$, the central energy of the most intense S$_{3}$ peak. Our intent was not to assign unequivocally the phonons involved but just to show that the experimental trend and the thermal one differs strongly in the measured temperature range. In Figure S4 of the Supporting Materials we plot the ratio (n$_{1}$ + 1)(n$_{2}$ + 1)/(n$_{A1g}$+1) for different energy values of the two phonon involved, always with the condition that $\hbar\omega_{1}$+$\hbar\omega_{2}$=596 cm$^{-1}$. The curves are barely distinguishable, meaning that the particular values used for the BE statistics, under the aforementioned conditions, do not affect the analysis.

The \textit{D} feature vanishes at low temperature, as expected by the temperature-dependence of the BE ratio for phonons originating from a difference of two modes (see SM, section D). In fact, at low temperatures there are no phonons to be taken from the system or, in other words, the system cannot be cooled by phonon absorption.

The hampering of the S- Raman modes cannot be described with the reduction of the phonon occupancy only but also with the thermally-driven increase of the band gap that drives our excitation energy out-of-resonance (see Figure \ref{Temp_PL}b for the S3-mode and SM, section D for the other modes). We note that we neglect possible contribution of anharmonic effects in reducing the Raman peak intensity.

\section{Conclusions}
In conclusion, we have investigated Raman spectra of bulk MoSe$_{2}$ crystals using visible (1.96 and 2.33 eV) and near-infrared (1.16 eV) excitation energies. In the latter case, we have found a significant increase in the intensity of second-order Raman modes. We have attributed this enhancement to a Raman process resonant with the indirect band gap of the crystal. We have compared the second-order Raman spectra with the two-phonon density of states calculated using a DFT approach. We have found good agreement between the Raman spectra measured with 2.33 eV excitation energy and the two-phonon density of states integrated over the entire FBZ. This demonstrates that i) most of the high order features can be ascribed to two-phonon resonances and that ii) at this photon energy the second-order Raman spectrum is not resonant with any specific indirect electronic transition with momentum away from zero. Differently, by using near-infrared excitation energy, the two-phonon Raman spectrum shows a qualitative agreement with the two-phonon density of states restricted to \textbf{q*}, i.e. the wavevector of the indirect electronic transition, further indicating that this process is resonant with the MoSe$_{2}$ indirect band gap. To further corroborate this result, we have measured Raman spectra as a function of the sample temperature. By comparing the intensity variation of the Raman modes with the thermal statistic of phononic occupation, we have ascribed the peak intensity decrease to a thermally-driven out-of-resonance condition due to the increase of the band-gap energy at low temperatures.

Understanding Raman scattering is of primary importance for different applications, such as growth quality control or individuation of MoSe$_{2}$ minerals, and gives a precious insight on the phonon dispersion and its coupling with electrons at the bottom of the conduction band \cite{macheda2018magnetotransport}. This study broadens the understanding of the resonance Raman spectrum in MoSe$_2$ and shows that Raman scattering with near-infrared excitation energy is a precious complementary tool for the study of van der Waals semiconductors. In perspective, we envision that one could make use of a near-infrared tunable lasers to probe the phonon dispersion away from zone center by means of resonance Raman in small-gap semiconductors.

\begin{acknowledgments}
The authors thank Lorenzo Graziotto and Francesco Mauri for helpful
and friendly discussions. We acknowledge the European Union’s Horizon 2020 research and innovation program under grant agreements no. 881603-Graphene Core3. We acknowledge support from the PRIN2017 grant number 2017Z8TS5B. We acknowledge that the results of this research have been
achieved using the DECI resource Mahti CSC based in
Finland at https://research.csc.fi/-/mahti with support
from the PRACE aisbl. We also aknowledge PRACE for awarding us access to Juliot-Curie at TGCC, France.
\end{acknowledgments}

\newpage
%\bibliographystyle{unsrt}
%\bibliography{bib2D}
%\end{document}

%\begin{comment}


\end{document}